\documentclass[12pt]{article}
\usepackage{geometry}
\usepackage{graphicx}
\usepackage[FIGTOPCAP,nooneline]{subfigure}
\usepackage{bm}
\usepackage{amsmath, amsfonts, amssymb}
\usepackage{cite}
\usepackage{multirow}
\usepackage{afterpage,array,rotating}
\usepackage{url}
\usepackage{authblk}
\usepackage{color}
\usepackage{fontspec}
\usepackage{float}
\usepackage[normalem]{ulem} 

\defaultfontfeatures{Ligatures=TeX}
\usepackage{lipsum}

\usepackage[para,online,flushleft]{threeparttable}
\usepackage{multirow}
\usepackage{array}

\definecolor{red}{RGB}{255,99,71}
\definecolor{blue}{RGB}{65,105,225}

\newcommand{\jkdone}[1]{\textcolor{magenta}{\bf [Jiyoung: done]}}
\newcommand{\jkprogress}[1]{\textcolor{magenta}{\bf [Jiyoung: in progress]}}

\definecolor{forestgreen}{rgb}{0.33,0.61,0.34}

\title{Quantifying concurrency in event-based temporal network and hypergraph data}

\author[1]{Jiyoung Kang}
\author[2]{Hang-Hyun Jo}
\author[3, 4, 5, *]{Naoki Masuda}

\affil[1]{Department of Scientific Computing, Pukyong National University, Busan, Republic of Korea}
\affil[2]{Department of Physics, The Catholic University of Korea, Bucheon, Republic of Korea}
\affil[3]{Gilbert S.\ Omenn Department of Computational Medicine and Bioinformatics, University of Michigan, Ann Arbor, MI, USA}
\affil[4]{Department of Mathematics, University of Michigan, Ann Arbor, MI, USA}
\affil[5]{Center for Computational Social Science, Kobe University, Kobe, Japan}
\affil[*]{Corresponding author. E-mail address: naokimas@umich.edu}

\begin{document}
\date{}
\maketitle

\begin{abstract}
Many social, biological, and technological systems are recorded as sequences of time-stamped interactions. In such systems, concurrency, i.e., the tendency for an individual to participate in multiple interactions approximately at the same time, can strongly affect processes such as epidemic or information spreading. However, concurrency measures for event-based temporal network data are not established. We introduce edge-event correlation (EEC), a simple and interpretable measure that quantifies how similarly two connections are active over time. We apply EEC to empirical temporal networks and temporal hypergraphs, the latter allowing single events to involve more than two nodes. Across most datasets, pairs of edges or hyperedges that share a node show higher concurrency than pairs that do not. We further find that this elevated concurrency is mainly driven by pairs embedded in closed local structures, such as triangles in the aggregated network. EEC provides a practical tool for quantifying concurrency in event-based temporal data and may help identify network structures that facilitate rapid spreading or collective dynamics.
\end{abstract}


\section{Introduction}

Empirical networks often vary over time on a timescale comparable to that of dynamical processes occurring on them. This observation has motivated studies of temporal (i.e., time-varying) networks for more than two decades, revealing key differences between temporal networks and their static counterparts\cite{holme2012temporal,Masuda2020Guide,z.Bansal2010Dynamic,z.Holme2023Temporal,z.Holme2015Modern}. Temporal network data are often provided as a list of time-stamped events between pairs of nodes. In other words, each edge of the network is associated with a sequence of events, each with an event time (and often an event duration). The simplest assumption for sequences of time-stamped events on edges, which implicitly underlies many ODE and agent-based models of dynamics on networks, is that these event sequences follow Poisson processes, in which events occur independently. However, empirical event sequences in networks often deviate from Poisson processes. Instead, they show burstiness, in which events on individual edges, or at individual nodes, cluster in time; this behavior is succinctly characterized by heavy-tailed distributions of inter-event times\cite{z.Barabasi2005Origin,z.Vazquez2006Modeling, KarsaiJoKaski2018}. Poisson processes have exponentially distributed inter-event times and are therefore considered non-bursty. Bursty event sequences have implications for dynamics on networks such as epidemic spreading~\cite{z.Vazquez2007Impact,z.Karsai2011Small,z.Masuda2013Predicting}.

In addition to burstiness, temporal organization of time-stamped events across edges may substantially affect dynamical processes in networked populations. Temporal motifs, which extend network motifs for static networks~\cite{milo2002networkmotifs,alon2007networkmotifs} to temporal networks, are a major tool for investigating such organization of time-stamped events.
A temporal motif is a small subgraph of time-stamped events whose participating nodes and edges, together with the temporal ordering of the events, jointly form a recurring pattern~\cite{Kovanen2011, Paranjape2017, JazayeriYang2020, Liu2023IeeeTransKnowlDataEng, z.Hosseinzadeh2022Temporal, chen2024tempme, Sariyuce2025_Lens}.
For example, consider three nodes $v_1$, $v_2$, and $v_3$, and two consecutive events on edges $(v_1, v_2)$ at time $t$ and $(v_2, v_3)$ at a later time $t'>t$, with $t'-t$ lying within a short time window. If, after aggregating across all node triples in the network, this temporally ordered $v_1\rightarrow v_2\rightarrow v_3$ pattern occurs more often than expected under a randomized null model, then the corresponding temporal motif is over-represented. Such over-representation may have dynamical consequences because, for example, this motif tends to make information or a pathogen propagate more quickly from $v_1$ to $v_3$ than from $v_3$ to $v_1$.

Concurrency of time-stamped events is a related type of temporal organization. The concept originates in mathematical and field epidemiology, where ``concurrent partnerships'' refer to a node (typically an individual) participating in two or more interaction relationships whose active intervals overlap or are interleaved in time~\cite{z.Morris1995Concurrent,z.Kretzschmar1996Measures,masuda2021concurrency}.
By extension, concurrency in a temporal network refers to the property that edges incident to a common node carry events whose timing coincides or clusters in close succession~\cite{masuda2021concurrency}. Concurrency is distinct from temporal motifs in that the former concerns whether events on different edges co-occur or cluster in time, irrespective of any specific temporal ordering between the events. Concurrency has been studied in mathematical and field epidemiology for decades, and it has long been recognized as a key factor in the epidemic potential of HIV~\cite{z.Kretzschmar1996Measures,WATTS199289, GoodreauCassels2012}.
Recent studies that separate the effect of concurrency from that of the underlying network structure have shown that concurrency promotes epidemic spreading on temporal networks~\cite{MoodyBenton2016, onaga2017concurrency, MillerSlim2017, LeeEmmonsGibsonMoodyMucha2019, BauchRand2000, Liu2024EurJApplMath}.

However, quantification of concurrency in time-stamped interaction data has lagged behind. For time-stamped events that last for some time (e.g., sexual partnerships), measures of concurrency have been developed~\cite{Lagarde2001, masuda2021concurrency}. However, many time-stamped interaction datasets that have become increasingly available in temporal network studies and adjacent fields are provided as sequences of instantaneous events. Although one can infer an effective duration of such events as the number of consecutive time windows in which two nodes have instantaneous time-stamped events, it is unclear how existing concurrency measures apply to such data. Given the aforementioned evidence that concurrency of time-stamped events in networks can promote epidemic spreading, and the abundance of time-stamped event data forming networks, it is imperative to develop clear and easily interpretable measures of concurrency for such data. In this study, we introduce a metric called the edge-event correlation (EEC) that quantifies event concurrency between edge pairs with time-stamped event sequences. We apply the EEC to empirical temporal networks and hypergraphs. In particular, we compare the EEC between (hyper)edge pairs that share a node (called intersecting pairs) and those that do not (called disjoint pairs), finding that intersecting pairs tend to have higher concurrency than disjoint pairs in empirical data. We further show that the enhanced EEC is mostly driven by intersecting pairs whose non-shared nodes are adjacent by a (hyper)edge closing a loop (see ``intersecting-closed pairs'' in Fig.~\ref{fig:schem}B).

\section{Methods}

\subsection{Edge-event correlation (EEC)}\label{sec:eec}

\begin{figure}
\centering
\includegraphics[width=0.7\textwidth]{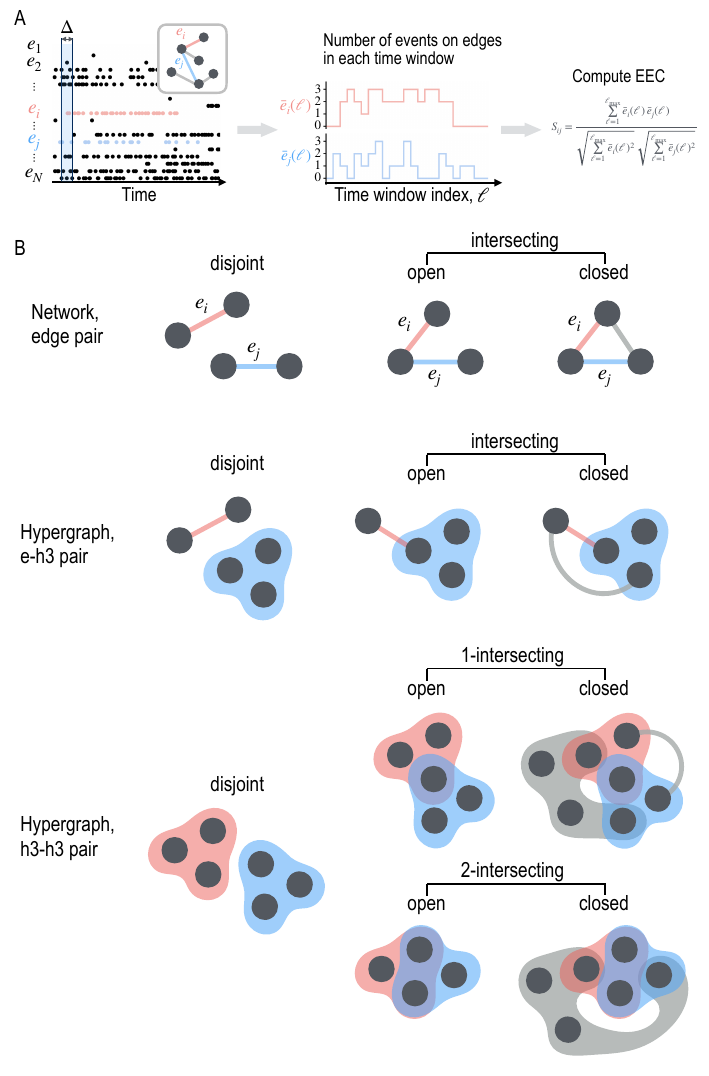}
\caption{Computation and comparison of the EEC. (A) Steps for computing the EEC. In the left panel, each dot represents a time-stamped event on an edge. For each edge, we count the number of events in each non-overlapping time window of size $\Delta$ to generate a sequence of event counts. Such sequences are shown in the middle panel for two edges that share a node (i.e., forming an intersecting edge pair). For each edge pair, which may be disjoint (i.e., non-intersecting) or intersecting, we compute the EEC. We then compare the EEC between groups of edge pairs, such as disjoint and intersecting pairs. (B) Different types of edge and hyperedge pairs in temporal networks and hypergraphs.}
\label{fig:schem}
\end{figure}

We define the edge-event correlation (EEC) as follows (Fig.~\ref{fig:schem}A).

We assume that time is discrete and represents events as occurring at integer time indices. Given a time-window width $\Delta$ (a positive integer measured in the native time unit of the data), we partition the observation period into non-overlapping windows of length $\Delta$. For the $i$th edge of the temporal network and the $\ell$th time window, we define
\begin{equation}
\bar e_i(\ell) = \#\{\,t : t \in [(\ell-1)\Delta,\, \ell\Delta) \text{ and a time-stamped event occurs on the $i$th edge at time } t\,\},
\label{eq:bar_e}
\end{equation}
where $\ell \in \{ 1, \ldots, \ell_{\max} \}$ and $\ell_{\max}$ is the largest integer such that the window $[(\ell_{\max}-1)\Delta,\, \ell_{\max}\Delta)$ is contained in the observation period. In other words, $\bar e_i(\ell)$ is the number of time-stamped events that occur on the $i$th edge in the $\ell$th time window.

To compute the EEC reliably, we restrict our attention to frequent edges, defined as those edges on which the total number of time-stamped events over the observation period is at least~$\theta$.
For a given $\Delta$ and any pair of frequent edges, the $i$th and $j$th edges, we define the edge-event correlation $S_{ij}$ as the cosine similarity between their event-count time series:
\begin{equation}
S_{ij} = \frac{\sum_{\ell=1}^{\ell_{\max}} \bar e_i(\ell)\,\bar e_j(\ell)}{\sqrt{\sum_{\ell=1}^{\ell_{\max}} \bar e_i(\ell)^{2}}\,\sqrt{\sum_{\ell=1}^{\ell_{\max}} \bar e_j(\ell)^{2}}}.
\label{eq:eec}
\end{equation}
Because $\bar e_i(\ell)\geq 0$, the EEC defined by Eq.~\eqref{eq:eec} takes values in $[0, 1]$. Two edges whose events occupy exactly the same windows (and in proportional counts) have $S_{ij}=1$. Two edges that produce events in disjoint windows have $S_{ij}=0$.
A similar cosine-similarity-based metric was proposed in~\cite{Schreiber2003} for measuring spike-timing reliability of a neuron in response to repeated presentation of a stimulus.

We have also verified that our main results remain qualitatively the same when we use two alternative definitions of the EEC (see Appendix~\ref{sec:app_pearson} for results). One is the Pearson correlation coefficient between the vectors $\{ \bar e_i(\ell) \}$ and $\{ \bar e_j(\ell) \}$. The other is the earth mover's distance between the original sequences of time-stamped events~\cite{SihnKim2019}.

\subsection{Comparing EEC between different types of edge pairs}

We compare the EEC between different types of edge pairs. Our primary comparison is between edge pairs that do not share any node (i.e., disjoint pairs) and those that share a node (i.e., intersecting pairs); see Fig.~\ref{fig:schem} for a schematic.

For two types of edge pairs, such as disjoint and intersecting edge pairs, we compare the mean EEC between the two edge-pair types. A standard statistical quantity would be the two-sided $p$-value (with unequal variances) from Welch's unpaired $t$-test. However, the number of samples in each edge-pair group differs by orders of magnitude across datasets, and the $p$-value tends to be driven mostly by sample size. In particular, because the number of edge pairs is large in many datasets, the $p$-value can be exceedingly small even when the difference in means is modest; this is a common statistical issue. Therefore, we instead measure Cohen's effect size $d=(\mu_{\rm int}-\mu_{\rm dj})\big/\sqrt{(\sigma^{2}_{\rm int}+\sigma^{2}_{\rm dj})/2}$ for each comparison~\cite{cohen1988}. Here, $\mu_{\rm dj}$ and $\mu_{\rm int}$ are the EEC values averaged over all edge pairs in the disjoint and intersecting edge-pair groups, respectively; $\sigma_{\rm dj}$ and $\sigma_{\rm int}$ are the corresponding standard deviations. Cohen's $d$ is interpreted on a domain-independent scale (small $\approx 0.2$, medium $\approx 0.5$, large $\approx 0.8$)~\cite{cohen1988}.

\section{Concurrency in empirical temporal networks and hypergraphs}\label{sec:results}

\subsection{Temporal networks}\label{sec:results_undirected}

\subsubsection{Data}\label{sec:results_undirected_data}

We surveyed ten general-purpose network data repositories and applied four inclusion criteria, retaining five genuinely pairwise temporal networks for the analysis below (shown in the first five rows in Table~\ref{tbl:dataset_undirected}). The full selection procedure is described in Appendix~\ref{sec:app_data_selection}.
We also used empirical temporal-hypergraph data to extract temporal networks. In particular, all 17 temporal hypergraphs are taken from the dataset collection accompanying~\cite{benson2018simplicial}, which is a common pool in the contemporary higher-order temporal-network literature.
We show the dyadic slices of these hypergraphs in the last 17 rows in Table~\ref{tbl:dataset_undirected}.
For each temporal hypergraph, we retain only the size-two (dyadic) hyperedges in this section; we analyze temporal hyperedges of size three in section~\ref{sec:results_hyper}.
Together, these 22 datasets span four orders of magnitude in node count and the length of the observation period, with time resolutions ranging from one second to one year.

For the EEC analysis, we keep only frequent edges, defined as edges with at least $\theta = 50$ time-stamped events. We require that the number of distinct frequent dyadic edges, denoted by $E_{\rm freq}$, is at least $50$.
We also require that the number of intersecting edge pairs, denoted by $n_{\rm int}$, and the number of disjoint edge pairs, denoted by $n_{\rm dj}$, are both at least $20$, because we statistically compare the EEC between disjoint and intersecting edge pairs.
Among the 22 temporal networks shown in Table~\ref{tbl:dataset_undirected}, 11 temporal networks satisfy these conditions. We examine the EEC for these 11 temporal networks.

\begin{table}[H]
\caption{Twenty-two empirical temporal networks analyzed. $N$ is the number of nodes that participate in at least one dyadic event; $E$ is the number of distinct dyadic edges; $E_{\rm freq}$ is the number of frequent edges, i.e., dyadic edges with at least $\theta=50$ time-stamped events; ``time res.'' is the native sampling resolution; in this and the $T$ column, s, d, q, and y abbreviate second, day, quarter, and year, respectively; $T$ is the observation span; $n_{\rm dj}$ and $n_{\rm int}$ are the numbers of disjoint and intersecting pairs among the frequent edges. The first five rows are genuinely pairwise temporal networks. The remaining rows correspond to the dyadic slice of empirical temporal hypergraphs.}
\label{tbl:dataset_undirected}
\centering
\begin{tabular}{l r r r l l r r}
\hline
Dataset & $N$ & $E$ & $E_{\rm freq}$ & time res. & $T$ & $n_{\rm dj}$ & $n_{\rm int}$ \\
\hline
\texttt{ia-reality-call} & 6,809 & 7,680 & 186 & 1 s & 106~d & 16,803 & 402 \\
\texttt{copenhagen-calls} & 536 & 621 & 7 & 1 s & 28~d & 21 & 0 \\
\texttt{college-msg} & 1,899 & 13,838 & 72 & 1 s & 194~d & 2,458 & 98 \\
\texttt{friends-family-call} & 129 & 392 & 117 & 1 s & 504~d & 6,348 & 438 \\
\texttt{social-evolution-call} & 75 & 270 & 26 & 1 s & 295~d & 314 & 11 \\
\texttt{contact-high-school} & 327 & 5,498 & 551 & 20 s & 4.2~d & 149,341 & 2,184 \\
\texttt{contact-primary-school} & 242 & 7,748 & 414 & 20 s & 1.4~d & 83,939 & 1,552 \\
\texttt{DAWN} & 2,003 & 30,991 & 790 & 1 q & 7.8~y & 289,225 & 22,430 \\
\texttt{NDC-classes} & 565 & 297 & 86 & 1 d & 118~y & 3,654 & 1 \\
\texttt{NDC-substances} & 977 & 1,130 & 41 & 1 d & 118~y & 775 & 45 \\
\texttt{coauth-DBLP} & 683,676 & 693,363 & 16 & 1 y & 82~y & 120 & 0 \\
\texttt{coauth-MAG-Geology} & 328,716 & 275,736 & 4 & 1 y & 218~y & 6 & 0 \\
\texttt{coauth-MAG-History} & 239,998 & 160,885 & 74 & 1 y & 217~y & 2,695 & 6 \\
\texttt{congress-bills} & 1,642 & 13,871 & 56 & 1 d & 34~y & 1,520 & 20 \\
\texttt{email-Enron} & 142 & 809 & 34 & 1 d & 3.6~y & 518 & 43 \\
\texttt{email-Eu} & 945 & 12,753 & 688 & 1 s & 2.2~y & 233,039 & 3,289 \\
\texttt{tags-ask-ubuntu} & 2,722 & 28,138 & 98 & 1 h & 8.6~y & 4,334 & 419 \\
\texttt{tags-math-sx} & 1,511 & 25,253 & 868 & 1 h & 7.4~y & 360,453 & 15,825 \\
\texttt{tags-stack-overflow} & 40,832 & 399,051 & 8,126 & 1 h & 9.1~y & 32,035,787 & 976,088 \\
\texttt{threads-ask-ubuntu} & 72,346 & 88,301 & 0 & 1 h & 7.7~y & 0 & 0 \\
\texttt{threads-math-sx} & 112,306 & 319,601 & 3 & 1 h & 7.1~y & 3 & 0 \\
\texttt{threads-stack-overflow} & 1,836,700 & 5,210,916 & 5 & 1 h & 9.1~y & 9 & 1 \\
\hline
\end{tabular}

\end{table}

For each surviving dataset, we choose a window width $\Delta$ to be approximately $T/200$,
where $T$ is the length of the observation period in the native time unit. The exception is for datasets whose native resolution is already coarse (e.g.\ DAWN at quarter-year resolution), in which case we use $\Delta=1$. We show the chosen $\Delta$ value for each dataset in Table~\ref{tbl:undirected_djns}.

\subsubsection{Results}\label{sec:results_undirected_results}

\begin{figure}[t!]
\centering
\includegraphics[width=0.99\textwidth]{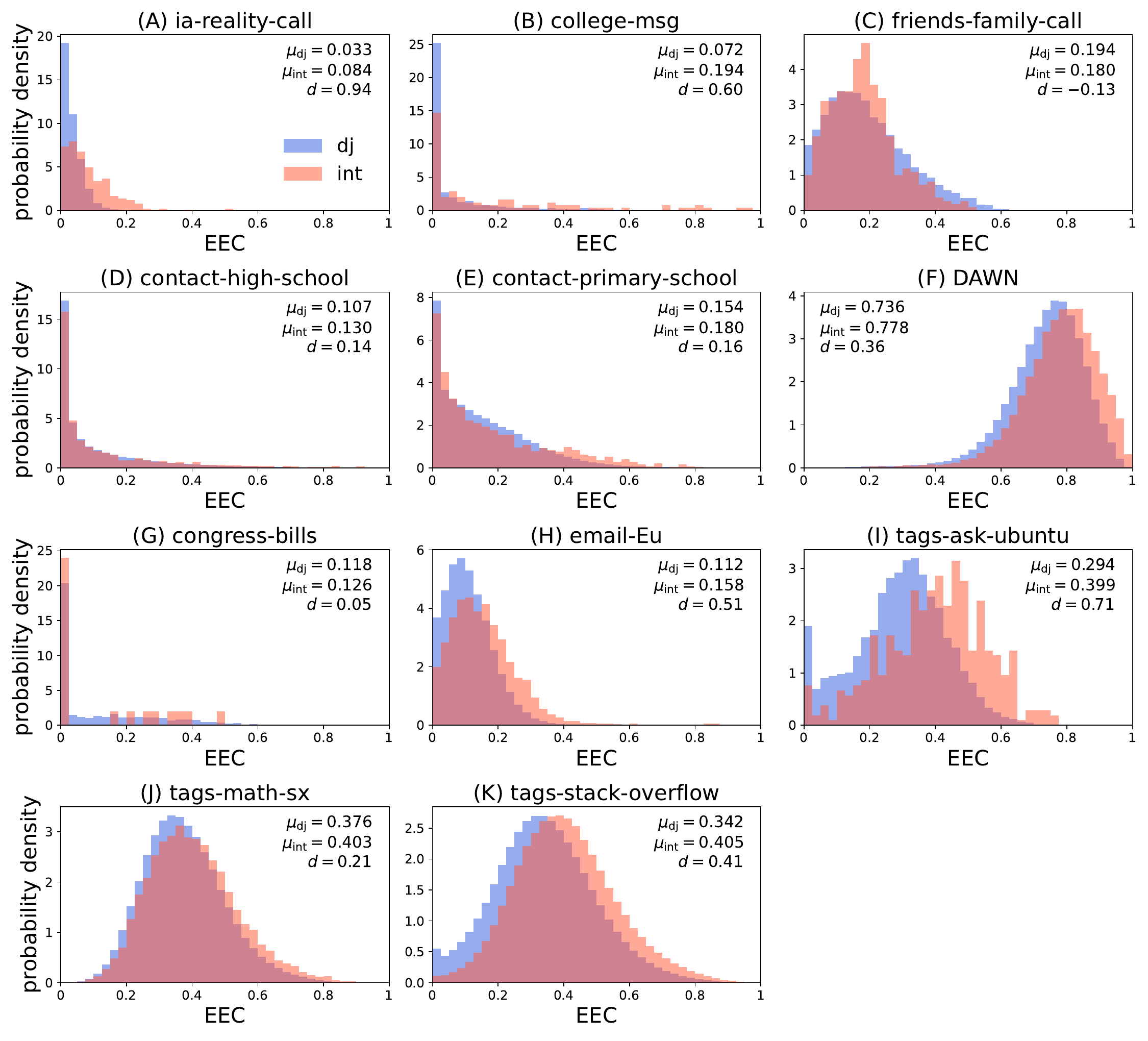}
\caption{Distribution of EEC values for different edge-pair types in temporal networks. Histograms show the probability density of EEC values for disjoint (blue) and intersecting (red) edge pairs in the 11 empirical datasets that satisfy $E_{\rm freq}\geq 50$, $n_{\rm dj}\geq 20$, and $n_{\rm int}\geq 20$. The mean EEC for each category is annotated in each panel, together with Cohen's $d$.}
\label{fig:fig2_undirected}
\end{figure}

Figure~\ref{fig:fig2_undirected} shows the distribution of EEC values for disjoint and intersecting edge pairs in the 11 datasets. We find that intersecting edge pairs have higher EEC values than disjoint edge pairs in ten of the eleven datasets. The exception is friends-family-call, in which the intersecting and disjoint means are close, with $d\approx -0.13$. Among the ten datasets in which $\mu_{\rm int}>\mu_{\rm dj}$, congress-bills shows the smallest effect ($d\approx 0.05$); this is consistent with the limited statistical power for that dataset, where $n_{\rm int}=20$.
Table~\ref{tbl:undirected_djns} as well as Fig.~\ref{fig:fig2_undirected} shows that, among the other nine datasets, the effect size $d$ ranges from very small (e.g., $\approx 0.14$--$0.16$)
to large ($\approx 0.94$).
Based on these results, we conclude that the EEC is overall higher for intersecting than for disjoint edge pairs.

Let us denote an intersecting edge pair by $\{ (v, v_1), (v, v_2) \}$, where $v$ is the shared node and $v_1$ and $v_2$ are the unshared nodes. We hypothesize that the higher concurrency of intersecting edge pairs is at least partially driven by the existence of the third edge $(v_1, v_2)$. To test this hypothesis, we sub-classify intersecting edge pairs into the following two classes (see Fig.~\ref{fig:schem}B):
\begin{itemize}
\item intersecting-open (acronym $\mathrm{int}_{\rm open}$): $(v_1, v_2)$ is not an edge in the static network. In other words, there are no time-stamped events between $v_1$ and $v_2$.
\item intersecting-closed (acronym $\mathrm{int}_{\rm closed}$): $(v_1, v_2)$ is an edge in the static network. Therefore, $v$, $v_1$, and $v_2$ form a triangle.
\end{itemize}
For the three-class comparison, we additionally require, on top of $n_{\rm dj}\geq 20$, that each of the two intersecting subclasses contain at least 10 pairs (i.e., $n_{{\rm int}_{\rm open}}\geq 10$ and $n_{{\rm int}_{\rm closed}}\geq 10$). Among the 11 datasets analyzed in Fig.~\ref{fig:fig2_undirected}, ten (i.e., all except congress-bills) satisfy this criterion and are included in the three-class analysis. For the datasets that originate from temporal hypergraphs, we only consider the dyadic (size-two) hyperedges of the static aggregated network when checking whether $(v_1, v_2)$ is an edge: an intersecting pair is classified as intersecting-closed only if $(v_1, v_2)$ itself is a recorded dyadic edge in the data, regardless of whether $v_1$ and $v_2$ co-occur in any larger hyperedge.

\begin{figure}[t!]
\centering
\includegraphics[width=0.99\textwidth]{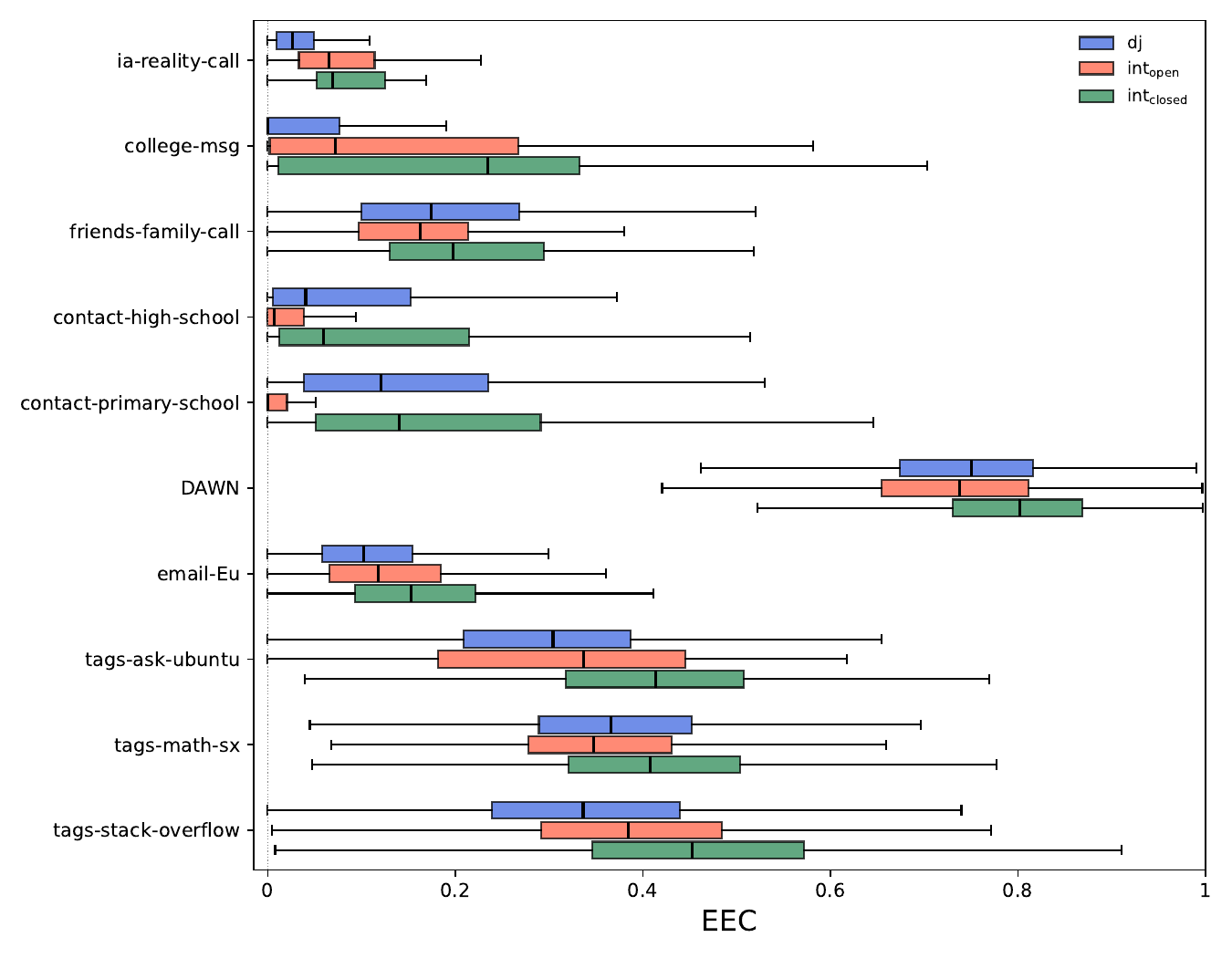}
\caption{Distribution of EEC values for the three edge-pair types (i.e., dj, $\mathrm{int}_{\rm open}$, and $\mathrm{int}_{\rm closed}$) in the ten temporal networks with sufficient samples. Each box spans the interquartile range (25th to 75th percentile), and the line inside the box indicates the median. Whiskers show the 5th and 95th percentiles; outliers are not shown.}
\label{fig:fig3_ns_split}
\end{figure}

We show the three-way comparison in terms of the EEC in Fig.~\ref{fig:fig3_ns_split} and the later columns of Table~\ref{tbl:undirected_djns}, which together show two patterns.
First, $\mu_{{\rm int}_{\rm closed}}>\mu_{\rm dj}$ in every temporal network, with an effect size $d_{\rm closed-dj}\geq 0.15$ in all networks.
Second, whether $\mu_{{\rm int}_{\rm open}}$ is larger or smaller than $\mu_{\rm dj}$ depends on the network: $d_{\rm open-dj}$ takes both positive and negative values across the ten networks, ranging from approximately $-1.4$ to $0.93$.
Intersecting-closed pairs are therefore the dominant contributors to the elevated concurrency seen in the disjoint-versus-intersecting comparison.
 We therefore conclude that the existence of the $(v_1, v_2)$ edge, together with sharing the node $v$, is the main driver of concurrency in temporal networks.

\begin{table}[H]
\caption{Comparison of EEC for edge pairs in the 11 temporal networks that satisfy $E_{\rm freq}\geq 50$, $n_{\rm dj}\geq 20$, and $n_{\rm int}\geq 20$. We denote by $\Delta$ the time-window width in native units; $\mu_{\rm dj}$, $\mu_{\rm int}$, $\mu_{\rm open}$, and $\mu_{\rm closed}$ are the mean EEC values in each category; $d$ is Cohen's effect size for the disjoint-versus-intersecting comparison; $d_{\rm open-dj}$ and $d_{\rm closed-dj}$ are Cohen's effect sizes for intersecting-open versus disjoint and intersecting-closed versus disjoint, respectively. The three-class columns ($\mu_{{\rm int}_{\rm open}}$, $\mu_{{\rm int}_{\rm closed}}$, $d_{\rm open-dj}$, $d_{\rm closed-dj}$) are dashed for congress-bills, which does not satisfy the sample-size requirement for the three-class analysis ($n_{{\rm int}_{\rm open}}\geq 10$ and $n_{{\rm int}_{\rm closed}}\geq 10$).}
\label{tbl:undirected_djns}
\centering
\small
\begin{tabular}{l r r r r r r r r r}
\hline
Dataset & $\Delta$ & $E_{\rm freq}$ & $\mu_{\rm dj}$ & $\mu_{\rm int}$ & $d$ & $\mu_{\rm open}$ & $\mu_{\rm closed}$ & $d_{\rm open{-}dj}$ & $d_{\rm closed{-}dj}$ \\
\hline
\texttt{ia-reality-call} & 3600 & 186 & 0.0330 & 0.0842 & 0.94 & 0.0832 & 0.0962 & 0.93 & 1.08 \\
\texttt{college-msg} & 86400 & 72 & 0.0716 & 0.1936 & 0.60 & 0.1835 & 0.2455 & 0.55 & 0.94 \\
\texttt{friends-family-call} & 86400 & 117 & 0.1940 & 0.1798 & $-0.13$ & 0.1630 & 0.2116 & $-0.29$ & 0.15 \\
\texttt{contact-high-school} & 90 & 551 & 0.1068 & 0.1302 & 0.14 & 0.0384 & 0.1462 & $-0.57$ & 0.23 \\
\texttt{contact-primary-school} & 30 & 414 & 0.1542 & 0.1803 & 0.16 & 0.0126 & 0.1932 & $-1.40$ & 0.25 \\
\texttt{DAWN} & 1 & 790 & 0.7362 & 0.7779 & 0.36 & 0.7174 & 0.7928 & $-0.15$ & 0.51 \\
\texttt{congress-bills} & 60 & 56 & 0.1184 & 0.1263 & 0.05 & -- & -- & -- & -- \\
\texttt{email-Eu} & 86400 & 688 & 0.1115 & 0.1583 & 0.51 & 0.1389 & 0.1672 & 0.29 & 0.62 \\
\texttt{tags-ask-ubuntu} & 400 & 98 & 0.2937 & 0.3987 & 0.71 & 0.3135 & 0.4048 & 0.12 & 0.77 \\
\texttt{tags-math-sx} & 325 & 868 & 0.3760 & 0.4028 & 0.21 & 0.3565 & 0.4192 & $-0.17$ & 0.33 \\
\texttt{tags-stack-overflow} & 400 & 8,126 & 0.3421 & 0.4053 & 0.41 & 0.3917 & 0.4643 & 0.33 & 0.76 \\
\hline
\end{tabular}

\end{table}

\subsection{Temporal hypergraphs}\label{sec:results_hyper}

\subsubsection{Data}\label{sec:results_hyper_data}

We now turn to potential concurrency of time-stamped events in the same temporal-hypergraph datasets as those considered in section~\ref{sec:results_undirected}, but now treat them as higher-order objects. Specifically, we keep the dyadic edges and add hyperedges of size three to the analysis. We neglect hyperedges of size larger than three because, in general, there are not sufficiently many frequent hyperedges of size larger than three to allow statistically reliable analyses. We start with the 17 temporal hypergraphs considered in section~\ref{sec:results_undirected_data}.

For size-three hyperedges, we apply a frequent-edge threshold of $\theta=25$. This is lower than the threshold $\theta=50$ used for edges in section~\ref{sec:results_undirected}. This choice is because size-three time-stamped events are generally rarer than size-two (i.e., edge) time-stamped events~\cite{benson2018simplicial, cencetti2021temporal, Ceria2025EpjDataSci, Mancastroppa2025PhysRevE, joMasuda2026hypergraph}.
For each temporal hypergraph, we use the same window size $\Delta$ for computing the EEC as that used in section~\ref{sec:results_undirected} (see Table~\ref{tbl:dataset_undirected} for the $\Delta$ values).

We apply the following dataset-selection criteria, which mirror those of section~\ref{sec:results_undirected_data}.

Consider a pair $(e_1, e_2)$ consisting of a frequent dyadic edge $e_1$ and a frequent size-three hyperedge $e_2$.
We label such a pair e-h3 (e for the dyadic edge and h3 for the size-three hyperedge). By definition, $e_1$ and $e_2$ in a disjoint e-h3 pair share no node, whereas those in an intersecting e-h3 pair share exactly one node; see Fig.~\ref{fig:schem}B. (We exclude pairs in which the dyadic edge lies inside the size-three hyperedge, i.e., shares two nodes.) We retain a temporal hypergraph for the disjoint-versus-intersecting analysis if the total number of frequent hyperedges across both sizes, $E_{2,{\rm freq}}+E_{3,{\rm freq}}$, is at least $40$ and $n_{\rm dj}\geq 20$ and $n_{\rm int}\geq 20$. For the three-class breakdown into intersecting-open ($\mathrm{int}_{\rm open}$) and intersecting-closed ($\mathrm{int}_{\rm closed}$), we further require $n_{{\rm int}_{\rm open}}\geq 10$ and $n_{{\rm int}_{\rm closed}}\geq 10$. The triangle-closure rule used to assign an e-h3 pair to $\mathrm{int}_{\rm closed}$ is a hypergraph-aware generalization of the size-two case: an e-h3 pair $(e_1, e_2)$ with one shared node is in $\mathrm{int}_{\rm closed}$ if at least one node pair $(v_i, v_j)$ with $v_i\in e_1\setminus(e_1\cap e_2)$ and $v_j\in e_2\setminus(e_1\cap e_2)$ co-occurs in a static hyperedge of any size; otherwise, the e-h3 pair is in $\mathrm{int}_{\rm open}$.

Consider a pair $(e_1, e_2)$ of size-three hyperedges, which we call an h3-h3 pair.
Two hyperedges forming an h3-h3 pair may share zero, one, or two nodes. (Sharing all three nodes would make them the same hyperedge and is excluded.) We split intersecting h3-h3 pairs into two types (see Fig.~\ref{fig:schem}B). An intersecting h3-h3 pair is said to be 1-intersecting (acronym $\mathrm{int}_1$) if $e_1$ and $e_2$ share exactly one node.
In contrast, the pair is 2-intersecting (acronym $\mathrm{int}_2$) if $e_1$ and $e_2$ share exactly two nodes.
To compare disjoint, 1-intersecting, and 2-intersecting h3-h3 pairs, we retain a temporal hypergraph if $E_{3,{\rm freq}}\geq 40$, $n_{\rm dj}\geq 20$, $n_{{\rm int}_1}\geq 20$, and $n_{{\rm int}_2}\geq 20$, where $n_{{\rm int}_1}$ and $n_{{\rm int}_2}$ are the numbers of 1-intersecting and 2-intersecting h3-h3 pairs, respectively.

We further split the 1-intersecting and 2-intersecting groups into ``open'' and ``closed'' subclasses using the same triangle-closure criterion as above. Concretely, for 1-intersecting pairs, there are four cross-pairs $(v_i, v_j)$ between the unshared nodes of $e_1$ and $e_2$. The pair is in $\mathrm{int}_{1,{\rm closed}}$ if at least one such cross-pair co-occurs in a static hyperedge of any size; otherwise, it is in $\mathrm{int}_{1,{\rm open}}$ (see Fig.~\ref{fig:schem}B). For 2-intersecting pairs, there is exactly one such cross-pair, and the same rule yields $\mathrm{int}_{2,{\rm closed}}$ versus $\mathrm{int}_{2,{\rm open}}$. For the resulting five-class comparison (disjoint, $\mathrm{int}_{1,{\rm open}}$, $\mathrm{int}_{1,{\rm closed}}$, $\mathrm{int}_{2,{\rm open}}$, $\mathrm{int}_{2,{\rm closed}}$), we additionally require each of the four intersecting subclasses to contain at least ten h3-h3 pairs.

\subsubsection{Results}\label{sec:results_hyper_results}

We first compare disjoint e-h3 pairs and intersecting e-h3 pairs in terms of the EEC. The means and Cohen's $d$ values for temporal hypergraphs with sufficient samples (see section~\ref{sec:results_hyper_data} for the criteria) are listed in Table~\ref{tbl:hyper_2x3_djns}. We find that the EEC for intersecting pairs is greater than that for disjoint pairs in most hypergraphs. We then compare the EEC among disjoint, intersecting-open, and intersecting-closed e-h3 pairs. The results are also shown in Table~\ref{tbl:hyper_2x3_djns}. We excluded the tags-ask-ubuntu hypergraph from this analysis because there are not sufficiently many intersecting-open and intersecting-closed samples.
The patterns mirror those of the network case in section~\ref{sec:results_undirected_results}: First, intersecting-closed e-h3 pairs consistently form the most concurrent class in seven of the eight hypergraphs (i.e., all except contact-primary-school). Second, the higher concurrency of intersecting than disjoint e-h3 pairs is primarily driven by the large EEC of intersecting-closed pairs; Table~\ref{tbl:hyper_2x3_djns} indicates that the EEC is larger for intersecting-open than for disjoint pairs in some hypergraphs, and vice versa in others.

\begin{table}[H]
\caption{Comparison of EEC for e-h3 pairs in empirical temporal hypergraphs. Columns $\mu_{\rm dj}$, $\mu_{\rm int}$, $\mu_{\rm open}$, and $\mu_{\rm closed}$ are the mean EEC values for disjoint, intersecting (open and closed combined), intersecting-open, and intersecting-closed pairs, respectively; $d$ is Cohen's effect size for the disjoint-versus-intersecting comparison; $d_{\rm open-dj}$ and $d_{\rm closed-dj}$ are Cohen's effect sizes for intersecting-open versus disjoint and intersecting-closed versus disjoint, respectively. We excluded tags-ask-ubuntu from the intersecting-open and intersecting-closed analysis because of the small sample size.}
\label{tbl:hyper_2x3_djns}
\centering
\begin{tabular}{l r r r r r r r}
\hline
Dataset & $\mu_{\rm dj}$ & $\mu_{\rm int}$ & $d$ & $\mu_{\rm open}$ & $\mu_{\rm closed}$ & $d_{\rm open{-}dj}$ & $d_{\rm closed{-}dj}$ \\
\hline
\texttt{contact-high-school} & 0.0721 & 0.0855 & 0.09 & 0.0142 & 0.0915 & $-0.61$ & 0.13 \\
\texttt{contact-primary-school} & 0.0745 & 0.0398 & $-0.26$ & 0.0010 & 0.0425 & $-0.69$ & $-0.24$ \\
\texttt{DAWN} & 0.6642 & 0.6974 & 0.26 & 0.5905 & 0.6976 & $-0.52$ & 0.26 \\
\texttt{NDC-substances} & 0.3516 & 0.4695 & 0.61 & 0.4252 & 0.4857 & 0.40 & 0.69 \\
\texttt{email-Eu} & 0.0881 & 0.1049 & 0.23 & 0.0747 & 0.1068 & $-0.20$ & 0.26 \\
\texttt{tags-ask-ubuntu} & 0.2473 & 0.2983 & 0.38 & -- & -- & -- & -- \\
\texttt{tags-math-sx} & 0.2918 & 0.3093 & 0.15 & 0.2490 & 0.3095 & $-0.40$ & 0.15 \\
\texttt{tags-stack-overflow} & 0.2753 & 0.3124 & 0.28 & 0.2630 & 0.3178 & $-0.10$ & 0.31 \\
\hline
\end{tabular}

\end{table}

We now turn to h3-h3 pairs and first compare disjoint, 1-intersecting, and 2-intersecting pairs. For temporal hypergraphs with sufficiently many qualified h3-h3 pairs, Table~\ref{tbl:hyper_3x3_main} reports the per-class means of the EEC and Cohen's $d$ for the comparisons between 1-intersecting and disjoint pairs and between 2-intersecting and 1-intersecting pairs. We find that the EEC of 2-intersecting pairs is consistently higher than that of 1-intersecting pairs, with Cohen's $d$ (denoted by $d_{{\rm int}_2-{\rm int}_1}$ in Table~\ref{tbl:hyper_3x3_main}) ranging from approximately $0.20$ to $0.40$ across the six hypergraphs. By contrast, 1-intersecting pairs are not appreciably more concurrent than disjoint pairs in most datasets: $d_{{\rm int}_1-{\rm dj}} \le 0.18$ in five of the six hypergraphs. Therefore, we conclude that the elevated concurrency of intersecting h3-h3 pairs is driven primarily by 2-intersecting rather than 1-intersecting h3-h3 pairs.

\begin{figure}[t!]
\centering
\includegraphics[width=0.95\textwidth]{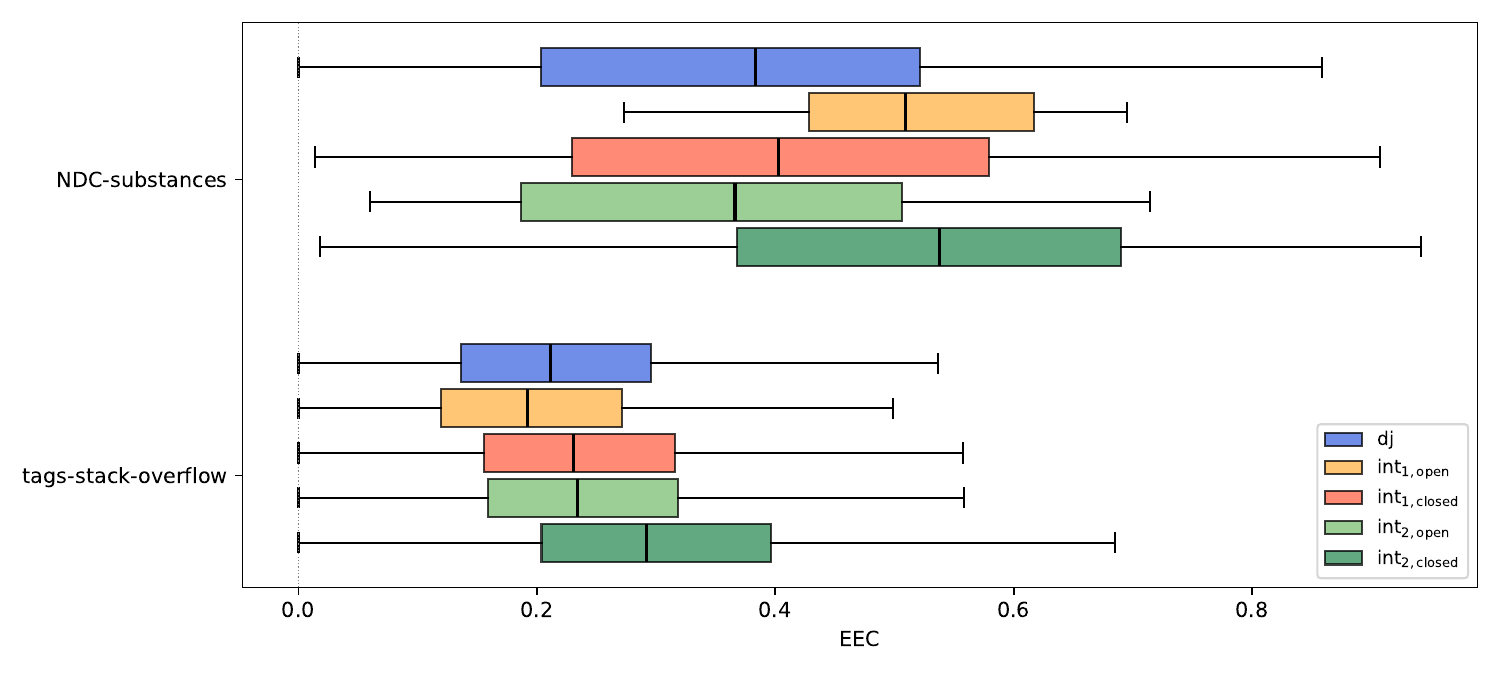}
\caption{Distribution of EEC values for the five h3-h3 pair types in empirical hypergraphs.}
\label{fig:fig_3x3_threeclass}
\end{figure}

Only two temporal hypergraphs contain sufficiently many h3-h3 pairs in each of the five subclasses (i.e., disjoint, $\mathrm{int}_{1,{\rm open}}$, $\mathrm{int}_{1,{\rm closed}}$, $\mathrm{int}_{2,{\rm open}}$, and $\mathrm{int}_{2,{\rm closed}}$). We compare the EEC across these subclasses of h3-h3 pairs in Fig.~\ref{fig:fig_3x3_threeclass}; we omit a table of $d$ values because it would require many pairwise comparisons between subclasses. Figure~\ref{fig:fig_3x3_threeclass} confirms a trend similar to that for edge pairs and e-h3 pairs. Specifically, h3-h3 pairs tend to be more concurrent when they share two nodes and the triangle is closed. This latter tendency does not hold for 1-intersecting h3-h3 pairs in the NDC-substances hypergraph; we do not have an explanation for this result.
We caution that this five-class breakdown is sample-limited, with only two temporal hypergraphs providing enough h3-h3 pairs in every subclass for a statistically viable comparison.

\begin{table}[H]
\caption{Comparison of EEC for disjoint, 1-intersecting (denoted by $\mathrm{int}_1$), and 2-intersecting (denoted by $\mathrm{int}_2$) types of h3-h3 pairs in empirical hypergraphs. Here, $\mu$ is the mean EEC, and $d$ is the effect size, with subscripts indicating the compared pair types.}
\label{tbl:hyper_3x3_main}
\centering
\begin{tabular}{l r r r r r}
\hline
Dataset & $\mu_{\rm dj}$ & $\mu_{{\rm int}_1}$ & $\mu_{{\rm int}_2}$ & $d_{{{\rm int}_1}{-}{\rm dj}}$ & $d_{{{\rm int}_2}{-}{{\rm int}_1}}$ \\
\hline
\texttt{DAWN} & 0.5944 & 0.6193 & 0.6671 & 0.18 & 0.33 \\
\texttt{NDC-substances} & 0.3702 & 0.4279 & 0.4831 & 0.27 & 0.24 \\
\texttt{email-Eu} & 0.0709 & 0.0677 & 0.0958 & $-0.05$ & 0.39 \\
\texttt{tags-ask-ubuntu} & 0.2190 & 0.2384 & 0.2618 & 0.16 & 0.20 \\
\texttt{tags-math-sx} & 0.2279 & 0.2400 & 0.2774 & 0.11 & 0.32 \\
\texttt{tags-stack-overflow} & 0.2229 & 0.2425 & 0.2958 & 0.16 & 0.40 \\
\hline
\end{tabular}

\end{table}

\section{Discussion}\label{sec:discussion}

In this study, we introduced the EEC to quantify concurrency in temporal networks. In terms of the EEC, we showed that various empirical temporal networks and hypergraphs have higher concurrency when two edges or hyperedges (we only say ``edges'' in the remainder of this section, but most of the following discussion also applies to hyperedges) share a node (i.e., intersecting pairs) than when they do not (i.e., disjoint pairs). This result is consistent with a node-activity view of temporal networks and hypergraphs, in which a node $v$ tends to generate time-stamped events with multiple neighbors when activated and tends to be silent when deactivated, thus promoting concurrency around $v$~\cite{Malmgren2008PNAS, Perra2012SciRep, Fonsecadosreis2020PhysRevE, Liu2024EurJApplMath, Hartle2025PhysRevResearch, joMasuda2026hypergraph}. However, the present data analysis provides only moderate (albeit consistent) support for larger EEC values for intersecting than for disjoint pairs; the effect size $d$ is modest in most comparisons.

Our second main result is that intersecting pairs that close a triangle (i.e., intersecting-closed pairs; see Fig.~\ref{fig:schem}B for a schematic) tend to have higher concurrency than intersecting pairs that do not (i.e., intersecting-open pairs), and that the former are the main driver of higher concurrency for intersecting pairs relative to disjoint pairs. This result is intuitive in that the three nodes forming a closed triangle are likely to be similar to each other in ways beyond their structural positions in networks~\cite{Granovetter1973, Asikainen2020, Peixoto2022}.
For example, in an intersecting edge pair $\{ (v, v_1), (v, v_2) \}$ in a co-authorship network, if $v_1$ and $v_2$ form an edge that closes the triangle, then the three researchers, $v$, $v_1$, and $v_2$, may share a research topic. Then, if $v$ works intensively on that research topic in a certain year, $v$ may coauthor one paper with $v_1$ and another with $v_2$ in the same year, raising concurrency. In contrast, if $v_1$ and $v_2$ are not adjacent, it is relatively more likely than in the closed-triangle case that a paper coauthored by $v$ and $v_1$ and one coauthored by $v$ and $v_2$ are on different research topics. In this case, if $v$ focuses on only one research topic in a single year, edges $(v, v_1)$ and $(v, v_2)$ may not occur in the same year, reducing concurrency.

The current study is limited by the availability of empirical data. There are a relatively small number of datasets that have sufficiently many frequent edges (i.e., edges with at least $\theta$ time-stamped events). Furthermore, our analysis requires sufficiently many intersecting pairs of frequent edges. This problem is exacerbated for hypergraphs because empirical temporal hypergraphs tend to have progressively fewer hyperedges of larger sizes~\cite{benson2018simplicial, cencetti2021temporal, Ceria2025EpjDataSci, Mancastroppa2025PhysRevE, joMasuda2026hypergraph}. Future data collection efforts may mitigate this problem. Examining the EEC for model temporal networks may also be useful.
We also note that, if a hub node $v$ has disproportionately many (say, $k_{\text{f}}$) frequent edges, it contributes quadratically many intersecting pairs (precisely, $k_{\text{f}} (k_{\text{f}}-1)/2$ pairs). Then, the EEC for intersecting pairs sharing $v$ may dominate the statistics of the EEC for intersecting pairs. A method to mitigate this effect may be worth developing.

For directed temporal networks, there are further natural classifications of intersecting edge pairs, while the computation of the EEC remains unchanged. Depending on the orientation of the edges, one can distinguish among $\{v \gets v_1, v \gets v_2\}$, $\{v \to v_1, v \to v_2\}$, and $\{v \gets v_1, v \to v_2\}$. Such a classification is analogous to that for network motifs~\cite{milo2002networkmotifs,alon2007networkmotifs}, which have predominantly been applied to directed networks.
A particular type of intersecting edge pair may have higher concurrency than others in empirical directed temporal networks. We leave this topic for future work.

Temporal motifs characterize time-lagged relationships between events on intersecting edge pairs in temporal networks~\cite{Kovanen2011, Paranjape2017, JazayeriYang2020, Liu2023IeeeTransKnowlDataEng, z.Hosseinzadeh2022Temporal, chen2024tempme, Sariyuce2025_Lens}, with recent extensions to temporal hypergraphs~\cite{LeeShin2023, Lotito2022, Gallo2024}.
Therefore, temporal motifs suggest the potential existence of causal relationships between events and nodes, or efficient pathways for information and pathogens (e.g., $v_1 \to v_2$ and then $v_2 \to v_3$ slightly later). Concurrency is a simpler concept than temporal motifs; concurrency neglects the ordering of events. However, two simultaneous events on intersecting edges similarly suggest that contagion across more than two nodes may occur efficiently during the time window in which the intersecting edges are simultaneously active. Pursuing relationships between the EEC, other forms of concurrency, and temporal motifs, including mathematical relationships, warrants future work.

Other definitions of EEC are possible. We have verified robustness of our results when we use
the Pearson correlation coefficient and earth mover's distance as the EEC metric
(see Appendix~\ref{sec:app_pearson}). Spike distance measures developed in computational neuroscience since 1990s
provide repertoires of EEC metrics; in computational neuroscience, a main goal is to compare how similar two or more sequences of spike trains, either coming from multiple neurons or repetitive recordings of a single neuron, are similar to each other.
Note that a spike train is a sequence of time-stamped events.
The Victor--Purpura metric~\cite{victor1996,victor1997} defines the distance between two spike trains as the minimum cost of an edit sequence that transforms one into the other, where each spike insertion or deletion costs one and shifting a spike by $\Delta t$ in time costs $q\,|\Delta t|$, with $q$ a parameter.
The van Rossum distance~\cite{vanrossum2001} convolves each spike train with a one-sided exponential decay function to produce a continuous signal, and then defines the distance between two spike trains as the $L^2$ norm of the difference between the two continuous signals.
We did not use these famous methods because they are rate-sensitive, i.e., sensitive
to differences in the total event count between two event sequences.
This property may not be a serious problem in neuroscience because one usually compares either spike trains from multiple neurons of similar types to assess the level of synchrony, or spike trains from the same neuron over multiple recording sessions to assess reproducibility of neuronal firing.
In contrast, in concurrency analysis on temporal networks, we compare event sequences on different edges, and different edges in a given network can have highly heterogeneous event rates. It should be noted that our EEC based on the cosine similarity as well as the alternatives considered in Appendix~\ref{sec:app_pearson} normalizes the event rate on each edge.

Other types of spike distance measures that may be useful as EEC for temporal networks include the event-synchronization family, which counts, for each spike in one train, the number of spikes in the other train that fall within an adaptive coincidence window \cite{QuianQuiroga2002}, and its time-resolved descendant SPIKE-Synchronization, which returns a binary coincidence indicator per spike — defined over the same adaptively chosen window — and reports the fraction of coincident spikes as the overall similarity \cite{Kreuz2013, Mulansky2015}. A related but methodologically distinct continuous-profile family — the ISI-distance, the SPIKE-distance, and their multi-scale extensions — quantifies dissimilarity through time-resolved profiles of locally defined inter-event-time or spike-timing mismatches between two event sequences \cite{Kreuz2007, Kreuz2013, Satuvuori2017}. Other related approaches include kernel-based multineuron metrics \cite{HoughtonSen2008}. See \cite{Victor2005, Kreuz2025} for reviews of these methods.
See \cite{Victor2005, Kreuz2025} for reviews of these methods.

In practical terms, the EEC can help identify pairs of edges, and the nodes incident to them, that contribute disproportionately to concurrency in a temporal network, beyond what would be expected from node degree alone. Such structures may be important for dynamical processes, including epidemic and information spreading, and may inform the design of targeted interventions. In conclusion, the EEC provides a simple and interpretable way to quantify concurrency in temporal networks and hypergraphs represented as sequences of time-stamped events. Our results show that concurrency is systematically elevated around intersecting pairs, especially when they participate in closed local structures, highlighting a link between temporal synchronization and local structural closure.

\section*{Acknowledgments}

JK acknowledges financial support by the National Research Foundation of Korea (NRF) grant funded by the Korea government (MSIT) (RS-2025-23323824). H-HJ acknowledges financial support by the National Research Foundation of Korea (NRF) grant funded by the Korea government (MSIT) (RS-2026-25476645). NM acknowledges financial support by the Japan Science and Technology Agency (JST) Moonshot R\&D (under Grant No. JPMJMS2021), the National Science Foundation (under grant no.\,2204936), and JSPS KAKENHI (under grant nos.\,JP 23H03414, 24K14840, and 24K03013).

During the preparation of this manuscript, the authors used Claude Opus 4.7 for language editing and code development assistance. All scientific interpretations, conclusions, code, and manuscript text were verified by the authors.

\bibliographystyle{unsrt}
\bibliography{ref_networkEEC}

\newpage
\appendix

\section{Selection of pairwise temporal networks}\label{sec:app_data_selection}

To assemble candidate temporal network datasets, we surveyed ten general-purpose network data repositories: Network Repository~\cite{NetworkRepository}, Netzschleuder~\cite{Netzschleuder}, the Koblenz Network Collection (KONECT)~\cite{KONECT}, the Stanford Network Analysis Platform (SNAP) dataset collection~\cite{SNAP}, the Colorado Index of Complex Networks (ICON)~\cite{ICON}, Reality Commons~\cite{RealityCommons}, Tore Opsahl's dataset~\cite{Opsahlweb}, the Temporal Graph Benchmark (TGB)~\cite{TGB}, the Dynamic Graph Benchmark (DGB)~\cite{DGB}, and SocioPatterns~\cite{SocioPatterns}. From each repository, we restricted attention to its temporal-network listings and applied four inclusion criteria.
First, we focused on undirected temporal networks; see section~\ref{sec:discussion} for discussion of directed temporal networks. The underlying interaction medium had to be inherently symmetric, such that the data are either published in undirected form or conventionally symmetrized in the literature. Second, we required that the data be genuinely pairwise networks. We excluded temporal networks in which simultaneously occurring three edges forming a triangle (i.e., on edges $(v_1, v_2)$, $(v_1, v_3)$, and $(v_2, v_3)$, where $v_1$, $v_2$, and $v_3$ are nodes) may in fact spuriously reflect a group interaction among the three nodes~\cite{benson2018simplicial, HaoLiu2026CommunPhys}. This filtering excludes face-to-face proximity and co-presence data (e.g., RFID/Bluetooth wearables, including the SocioPatterns datasets when treated as networks), multi-recipient e-mail, and online forum/thread interactions (e.g., Slashdot reply networks,
Wikipedia talk networks, and the Stack Overflow temporal network). These data are more faithfully modeled as temporal hypergraphs~\cite{benson2018simplicial,cencetti2021temporal}, which we examine in section~\ref{sec:results_hyper}. Third, we excluded bipartite event data, such as user--item or user--page interactions, because incorporating node types into the analysis is beyond the scope of this study. Fourth, we excluded gated, country-scale call-detail-record datasets~\cite{onnela2007structure,blondel2015survey} that are not publicly redistributable.

Applying these criteria removes the large majority of each repository's temporal-network listings. Nine datasets survive the filter, all of which are phone calls, SMS, or one-to-one instant messages.
The nine temporal networks are the calls and SMS layers of the MIT Reality Mining study~\cite{Eagle2006}, the Friends and Family study~\cite{Aharony2011}, the Social Evolution study~\cite{Madan2012}, and the Copenhagen Networks Study~\cite{Sapiezynski2019}; and the UC-Irvine online-message data~\cite{Opsahl2009}. Among these, the four SMS datasets (from the MIT Reality Mining, Friends and Family, Social Evolution, and Copenhagen Networks studies) are recorded directionally, with each record carrying a distinct sender and recipient. There is no convention in the literature to symmetrize SMS data into an undirected temporal network, so we exclude all four SMS datasets. The five remaining datasets are ia-reality-call (the MIT Reality Mining calls layer), copenhagen-calls (the calls layer of the Copenhagen Networks Study), college-msg (the UC-Irvine online messages), friends-family-call (the calls layer of the Friends and Family study), and social-evolution-call (the calls layer of the Social Evolution study). Phone calls between two people are conventionally treated as undirected because the medium itself is symmetric, and we follow this convention.

\section{Robustness tests for alternative definitions of the EEC}\label{sec:app_alt_eec}

In this section, we examine two alternative definitions of the EEC and check whether our main conclusions are preserved. We restrict the report to Cohen's effect size $d$ for each comparison because it provides the summary that supports the qualitative conclusions.

\subsection{EEC based on the Pearson correlation coefficient}\label{sec:app_pearson}

The first alternative definition replaces the cosine similarity with the Pearson correlation coefficient between the two event-count time series, namely,
\begin{equation}
r_{ij} = \frac{\sum_{\ell=1}^{\ell_{\max}} \bigl[\bar e_i(\ell) - \langle \bar e_i \rangle\bigr]\bigl[\bar e_j(\ell) - \langle \bar e_j \rangle\bigr]}{\sqrt{\sum_{\ell=1}^{\ell_{\max}} \bigl[\bar e_i(\ell) - \langle \bar e_i \rangle\bigr]^2}\,\sqrt{\sum_{\ell=1}^{\ell_{\max}} \bigl[\bar e_j(\ell) - \langle \bar e_j \rangle\bigr]^2}},
\label{eq:eec_pearson}
\end{equation}
where $\langle\bar e_i\rangle$ and $\langle\bar e_j\rangle$ are the per-edge means of the event-count series. This is the same expression as Eq.~\eqref{eq:eec} except that the per-edge mean is subtracted from each component before the inner product is taken. The Pearson EEC defined by Eq.~\eqref{eq:eec_pearson} takes values in $[-1, 1]$, in contrast to the cosine similarity in the main text, which takes values in $[0, 1]$.

\subsection{EEC based on the earth mover's distance}\label{sec:app_emd}

The second alternative is based on the spike-train distance proposed by Sihn and Kim~\cite{SihnKim2019}, which adapts the earth mover's distance (EMD). For distributions with equal total mass on the real line, the EMD coincides with the one-dimensional Wasserstein-1 distance, also called the Kantorovich--Rubinstein distance.
The EMD was popularized for similarity measurement in computer vision~\cite{RubnerTomasiGuibas2000} and is rooted in the classical Monge--Kantorovich optimal transport theory (see~\cite{Villani2009} for a comprehensive mathematical treatment); see~\cite{PeyreCuturi2019} for a computational review.
In contrast to the EEC defined by Eq.~\eqref{eq:eec} or its Pearson counterpart given by Eq.~\eqref{eq:eec_pearson}, the EMD operates directly on the raw event time stamps rather than on a binned event-count representation. Therefore, the EMD does not require the time-window width $\Delta$.

For the $i$th edge with $N_i$ time-stamped events at $\{x^{(i)}_1, \ldots, x^{(i)}_{N_i}\}$, we view the event sequence as a discrete probability distribution that places mass $1/N_i$ at each time stamp; the resulting distribution has total mass one regardless of $N_i$. The earth mover's distance between the $i$th and $j$th edges is given by
\begin{equation}
W_{ij} = \min_{\xi_{11}, \ldots, \xi_{N_i N_j}} \sum_{p=1}^{N_i} \sum_{q=1}^{N_j} \bigl|x^{(i)}_p - x^{(j)}_q\bigr|\, \xi_{pq},
\label{eq:eec_emd}
\end{equation}
subject to $\xi_{pq} \ge 0$ for $p \in \{1, \ldots, N_i \}$ and $q \in \{1, \ldots, N_j \}$, $\sum_{p=1}^{N_i} \xi_{pq} \leq 1/N_j$, $\sum_{q=1}^{N_j} \xi_{pq} \leq 1/N_i$, and $\sum_{p=1}^{N_i} \sum_{q=1}^{N_j} \xi_{pq} = 1$. The optimal $\{ \xi_{11}, \ldots, \xi_{N_i N_j} \}$ describes the minimum total amount of probability mass, weighted by the time distance over which it must be transported, that turns one event sequence into the other. Note that $W_{ij}$ is a distance rather than a similarity, so smaller $W_{ij}$ indicates higher concurrency between the two event sequences.

\subsection{Results}\label{sec:app_alt_results}

We repeat the per-pair-type comparisons of section~\ref{sec:results} using each of the two alternative definitions. We show the resulting Cohen's $d$ values in Tables~\ref{tbl:app_size2}, \ref{tbl:app_size2x3}, and~\ref{tbl:app_size3x3}, which correspond to Tables~\ref{tbl:undirected_djns}, \ref{tbl:hyper_2x3_djns}, and~\ref{tbl:hyper_3x3_main} of the main text, respectively.

The qualitative conclusions of the main text are preserved under both alternative definitions.
We recall that, for the earth mover's distance, the sign of $d$ has the opposite meaning from that for the cosine similarity and Pearson correlation coefficient. Specifically, for the earth mover's distance, $d < 0$ for an intersecting class versus the disjoint class, for example, indicates that the intersecting class is more concurrent than disjoint pairs.

\begin{table}[h!]
\caption{Cohen's $d$ for edge pairs under the Pearson correlation coefficient (PCC) and the earth mover's distance (EMD) as the EEC metrics. This table corresponds to Table~\ref{tbl:undirected_djns} of the main text. The three-class columns are dashed for congress-bills because it does not satisfy the sample-size requirement for the three-class analysis ($n_{{\rm int}_{\rm open}}\geq 10$ and $n_{{\rm int}_{\rm closed}}\geq 10$).}
\label{tbl:app_size2}
\centering
\small
\begin{tabular}{l r r r | r r r}
\hline
 & \multicolumn{3}{c|}{PCC} & \multicolumn{3}{c}{EMD} \\
Dataset & $d$ & $d_{\rm open{-}dj}$ & $d_{\rm closed{-}dj}$ & $d$ & $d_{\rm open{-}dj}$ & $d_{\rm closed{-}dj}$ \\
\hline
\texttt{ia-reality-call} & 0.97 & 0.96 & 1.11 & $-0.62$ & $-0.61$ & $-0.72$ \\
\texttt{college-msg} & 0.62 & 0.56 & 0.96 & $-0.74$ & $-0.79$ & $-0.50$ \\
\texttt{friends-family-call} & 0.00 & $-0.19$ & 0.32 & $-0.18$ & $-0.18$ & $-0.20$ \\
\texttt{contact-high-school} & 0.15 & $-0.51$ & 0.24 & $-0.08$ & 0.27 & $-0.13$ \\
\texttt{contact-primary-school} & 0.16 & $-1.25$ & 0.24 & $-0.16$ & 0.69 & $-0.22$ \\
\texttt{DAWN} & 0.16 & $-0.30$ & 0.27 & $-0.01$ & 0.46 & $-0.23$ \\
\texttt{congress-bills} & 0.04 & -- & -- & $-0.63$ & -- & -- \\
\texttt{email-Eu} & 0.53 & 0.34 & 0.64 & $-0.19$ & 0.06 & $-0.24$ \\
\texttt{tags-ask-ubuntu} & 0.84 & 0.21 & 0.89 & $-0.49$ & 2.17 & $-0.52$ \\
\texttt{tags-math-sx} & 0.20 & $-0.04$ & 0.28 & $-0.16$ & 0.25 & $-0.24$ \\
\texttt{tags-stack-overflow} & 0.30 & 0.26 & 0.49 & $-0.39$ & $-0.26$ & $-0.52$ \\
\hline
\end{tabular}

\end{table}

\begin{table}[h!]
\caption{Cohen's $d$ for e-h3 pairs under the PCC- and EMD-based EEC. This table corresponds to Table~\ref{tbl:hyper_2x3_djns} of the main text. The three-class columns are dashed for tags-ask-ubuntu because it does not satisfy the sample-size requirement for the three-class analysis.}
\label{tbl:app_size2x3}
\centering
\small
\begin{tabular}{l r r r | r r r}
\hline
 & \multicolumn{3}{c|}{PCC} & \multicolumn{3}{c}{EMD} \\
Dataset & $d$ & $d_{\rm open{-}dj}$ & $d_{\rm closed{-}dj}$ & $d$ & $d_{\rm open{-}dj}$ & $d_{\rm closed{-}dj}$ \\
\hline
\texttt{contact-high-school} & 0.11 & $-0.58$ & 0.15 & 0.01 & 0.25 & $-0.01$ \\
\texttt{contact-primary-school} & $-0.26$ & $-0.61$ & $-0.24$ & 0.06 & $-0.37$ & 0.09 \\
\texttt{DAWN} & 0.09 & $-0.44$ & 0.09 & $-0.09$ & 0.45 & $-0.10$ \\
\texttt{NDC-substances} & 0.61 & 0.38 & 0.69 & $-0.59$ & $-0.42$ & $-0.66$ \\
\texttt{email-Eu} & 0.24 & $-0.12$ & 0.26 & $-0.06$ & 0.28 & $-0.09$ \\
\texttt{tags-ask-ubuntu} & 0.44 & -- & -- & $-0.35$ & -- & -- \\
\texttt{tags-math-sx} & 0.14 & $-0.14$ & 0.14 & $-0.12$ & 0.21 & $-0.13$ \\
\texttt{tags-stack-overflow} & 0.19 & 0.05 & 0.20 & $-0.25$ & $-0.16$ & $-0.35$ \\
\hline
\end{tabular}

\end{table}

\begin{table}[h!]
\caption{Cohen's $d$ for h3-h3 pairs under the PCC- and EMD-based EEC. This table corresponds to Table~\ref{tbl:hyper_3x3_main}.}
\label{tbl:app_size3x3}
\centering
\begin{tabular}{l r r | r r}
\hline
 & \multicolumn{2}{c|}{PCC} & \multicolumn{2}{c}{EMD} \\
Dataset & $d_{{\rm int}_1{-}{\rm dj}}$ & $d_{{\rm int}_2{-}{\rm int}_1}$ & $d_{{\rm int}_1{-}{\rm dj}}$ & $d_{{\rm int}_2{-}{\rm int}_1}$ \\
\hline
\texttt{DAWN} & 0.05 & 0.28 & $-0.06$ & $-0.22$ \\
\texttt{NDC-substances} & 0.28 & 0.25 & $-0.73$ & 0.12 \\
\texttt{email-Eu} & $-0.03$ & 0.34 & 0.09 & $-0.28$ \\
\texttt{tags-ask-ubuntu} & 0.18 & 0.12 & $-0.26$ & $-0.24$ \\
\texttt{tags-math-sx} & 0.08 & 0.26 & $-0.14$ & $-0.03$ \\
\texttt{tags-stack-overflow} & 0.11 & 0.33 & $-0.20$ & $-0.37$ \\
\hline
\end{tabular}

\end{table}

\end{document}